\documentclass{sig-alternate-2013}
\usepackage{dsfont,resizegather,graphicx, multirow, hyperref, enumitem, flushend}

\newcommand\T{\rule{0pt}{2.6ex}}       
\newcommand\B{\rule[-1.2ex]{0pt}{0pt}}

\newfont{\mycrnotice}{ptmr8t at 7pt}
\newfont{\myconfname}{ptmri8t at 7pt}
\let\crnotice\mycrnotice%
\let\confname\myconfname%

\permission{Permission to make digital or hard copies of all or part of this work for personal or classroom use is granted without fee provided that copies are not made or distributed for profit or commercial advantage and that copies bear this notice and the full citation on the first page. Copyrights for components of this work owned by others than the author(s) must be honored. Abstracting with credit is permitted. To copy otherwise, or republish, to post on servers or to redistribute to lists, requires prior specific permission and/or a fee. Request permissions from Permissions@acm.org.}
\conferenceinfo{KDD'15,}{August 11-14, 2015, Sydney, NSW, Australia. \\
{\mycrnotice{Copyright is held by the owner/author(s). Publication rights licensed to ACM.}}}
\copyrightetc{ACM \the\acmcopyr}
\crdata{978-1-4503-3664-2/15/08\ ...\$15.00.\\
DOI: http://dx.doi.org/10.1145/2783258.2788570}

\begin{document}

\title{Predicting Ambulance Demand:\\a Spatio-Temporal Kernel Approach}

\numberofauthors{2} 
\author{
\alignauthor
Zhengyi Zhou \\
       \affaddr{Center for Applied Mathematics}\\
       \affaddr{Cornell University}\\
       \affaddr{Ithaca, NY 14853}\\
       \email{zz254@cornell.edu}
\alignauthor
David S. Matteson \\ 
       \affaddr{Department of Statistical Science}\\
       \affaddr{Cornell University}\\
       \affaddr{Ithaca, NY 14853}\\
       \email{matteson@cornell.edu}
}

\maketitle
\begin{abstract}
Predicting ambulance demand accurately at fine time and location scales is critical for ambulance fleet management and dynamic deployment. Large-scale datasets in this setting typically exhibit complex spatio-temporal dynamics and sparsity at high resolutions. We propose a predictive method using spatio-temporal kernel density estimation (stKDE) to address these challenges, and provide spatial density predictions for ambulance demand in Toronto, Canada as it varies over hourly intervals. Specifically, we weight the spatial kernel of each historical observation by its informativeness to the current predictive task. We construct spatio-temporal weight functions to incorporate various temporal and spatial patterns in ambulance demand, including location-specific seasonalities and short-term serial dependence. This allows us to draw out the most helpful historical data, and exploit spatio-temporal patterns in the data for accurate and fast predictions. We further provide efficient estimation and customizable prediction procedures. stKDE is easy to use and interpret by non-specialized personnel from the emergency medical service industry. It also has significantly higher statistical accuracy than the current industry practice, with a comparable amount of computational expense.
\end{abstract}



\section{Introduction}
The emergency medical service (EMS) industry aims to minimize response times to emergencies and keep operational costs low. Accurate spatio-temporal demand predictions are critical to staff / fleet management and dynamic deployment. Predictions are required at high temporal and spatial resolutions for operational decision-making; the industry typically predicts for every hour and every 1 km$^2$ region. We are motivated to predict spatio-temporal ambulance demand in Toronto, Canada.

There are several typical challenges to predicting ambulance demand. First, ambulance demand data is often sparse at the prediction resolution. For instance, Toronto receives only 23 highest priority events per hour on average; 96\% of the 1 km$^2$ spatial regions in Toronto have zero event in any hour. Also, ambulance demand frequently exhibits complex temporal dynamics, some of which are location-specific. For example, ambulance demand in Toronto has weekly and daily seasonality and short-term serial dependence of a few hours \cite{Matteson:2011, Zhou:2015a}. Among these patterns, daily seasonality and short-term dependence are more pronounced downtown and in denser neighborhoods \cite{Zhou:2015a}. Lastly, ambulance demand data in large cities is usually large-scale; Toronto dispatches for about 200,000 priority events per year. This may present computational challenges, especially since predictions are needed hourly. 

The current industry practice for predicting ambulance demand often uses a simple averaging formula. Demand in a 1 km$^2$ spatial region over an hour is typically predicted by averaging a small number of historical counts, from the same spatial region and over the corresponding hours from previous weeks or years \cite{Goldberg:2004}. In current practice, Toronto EMS averages four historical counts in the same hour of the year over the past four years, while the EMS of Charlotte-Mecklenburg, North Carolina averages twenty historical counts in the same hour of the preceding four weeks for the past five years (the MEDIC method) \cite{Setzler:2009}. Averaging so few historical counts, which are mostly zeros, produces highly noisy and flickering predictions, resulting in haphazard and inefficient deployment. Such methods are also sensitive to how the spatial domains are partitioned \cite{Goldberg:2004}.

Many studies have accurately predicted aggregate ambulance demand as a temporal process. These studies have considered autoregressive moving average models \cite{Channouf:2007}, factor models \cite{Matteson:2011} and spectrum analysis \cite{Vile:2012}. However, very few studies have modeled spatio-temporal ambulance demand well. One study uses artificial neural networks (ANN) to predict ambulance demand on discretized time and space, but fails to improve predictive accuracy over typical industry practice due to data sparsity \cite{Setzler:2009}. Another more recent study predicts ambulance demand in discrete time and continuous space \cite{Zhou:2015a}. It proposes a time-varying Gaussian mixture model (GMM) to incorporate location-specific temporal patterns in the demand, giving higher predictive accuracy than industry approaches. However, its estimation procedure via Bayesian sampling may present computational challenges and require special expertise, making it difficult to use by non-specialized personnel from the EMS industry. 

Kernel density estimation (KDE) is a powerful tool for non-parametric density estimation in spatio-temporal data. It has been widely applied to visualize or forecast spatio-temporal crime incidence \cite{Brunsdon:2007, Nakaya:2010}, disease spread \cite{Zhang:2011, Wilesmith:2003}, product demand \cite{Jansenberger:2004}, and data streams \cite{Aggarwal:2003, Procopiuc:2005}. In most cases, the time dimension is treated differently from the space dimension(s). The most traditional approach is to build a separate spatial KDE for each discrete time period. However, this approach may result in uneven subset size and sparse subsets with too little data for accurate density estimates. Recent studies assume a multiplicative orthogonal relationship between the time and space dimensions. For example, \cite{Aggarwal:2003} multiplies a spatial kernel with a linear function in time. Studies such as \cite{Brunsdon:2007, Nakaya:2010, Zhang:2011} multiply a spatial kernel and a temporal kernel with different bandwidths and kernel functions for the two kernels. 

Extending these multiplicative spatio-temporal KDE methods, we propose a fast and accurate method for predicting spatio-temporal ambulance demand that is practical and scalable in industrial settings. We follow \cite{Zhou:2015a} in predicting in discrete time and continuous space. We propose a novel specification of spatio-temporal kernel density estimation (stKDE). First, we learn parametrically the temporal and spatial characteristics of the demand. Each historical observation is annotated with a weight based on what we have learned. This spatio-temporal weight function scales how helpful different historical observations are to a given predictive task. Then we construct a spatial kernel density estimator weighted by the informativeness weight function, and use the resulting kernel density estimates as predictions. In this way, we efficiently emphasize the historical data most important to prediction and, as far as possible, exploit the spatial and temporal characteristics in the data. This approach has three main advantages
\begin{enumerate} [label=(\alph*)]
\item accessibility: stKDE is fully automated and robust. It is easy to interpret and use by non-specialized personnel, while approaches such as ANN or GMM may need special expertise (e.g., tuning, MCMC diagnostics). 
\item efficiency: stKDE has low computational complexity. It is faster than GMM; inferring latent component label in GMM is costly. It can afford more frequent parameter estimation updates and online predictions.
\item accuracy: stKDE gives more accurate predictions than current industry practice with similar computational expense. It also outperforms naive KDE methods (and ANN via \cite{Setzler:2009}); it is at least as accurate as GMM.
\end{enumerate} 

We implement this method on ambulance demand data from Toronto, Canada from years 2007 and 2008. The data consist of $391,296$ priority emergency events received by Toronto EMS for which an ambulance was dispatched. Each record contains the time and the location to which the ambulance was dispatched. This includes some calls not requiring lights-and-sirens response, but does not include scheduled patient transfers. We include only the first event in our analysis when multiple responses are received for the same event; explanatory analysis did not reveal any spatial or temporal pattern for these cases, and we treat them as a single ambulance dispatch. We have removed all calls with no times or locations. There was no call received for more than two hours on March 10, 2007 due to a recording system malfunction, and we have also removed all calls from that day. These removals totaled less than 2\% of the data. For example, Figure 1 shows the locations of all ambulance demand from January to July 2008. 

Our proposed method, stKDE, gives significantly more accurate predictions than current industry practice with similar computational expense. It also compares favorably to the state-of-the-art in ambulance demand prediction research. 

We propose the stKDE model in \S 2 and discuss computational methods in \S3. We show the empirical results on Toronto ambulance demand in Section \S 4, and conclude in \S 5.

\section{Methodology}
We model Toronto's ambulance demand on a continuous spatial domain $\mathcal{S}\subseteq \mathds{R}^2$  and a discretized temporal domain of one-hour intervals $\mathcal{T}=\{1,2, \ldots\}$. Let $\mathbf{s}_{t,i}$ be the location of the $i$-th ambulance demand arising from the $t$-th time period, for $i\in \{1,\ldots,n_t\}$, where $n_t$ is the total number of ambulances demanded in the $t$-th period. Since a non-homogeneous Poisson process (NHPP) is a natural model for a spatial point process \cite{Diggle:2003, Moller:2004}, we assume $\{\mathbf{s}_{t,i} : i = 1,\ldots n_t\}$ for each time period $t$ independently follow an NHPP over $\mathcal{S},$ with positive intensity function $\lambda_t$. We decompose intensity function as $\lambda_t(\mathbf{s})=\delta_t f_t(\mathbf{s})$, for $\mathbf{s}\in\mathcal{S}$. Here, $\delta_t=\int_{\mathcal{S}} \lambda_t(\mathbf{s})\, d\mathbf{s}$ is the aggregate demand intensity over the spatial domain, and $f_t(\cdot)$ is the continuous spatial \textit{density} of the demand at time $t$ such that $f_t(\mathbf{s})>0$ and $\int_\mathcal{S} f_t(\mathbf{s})d\mathbf{s}=1$.
Hence, for each $t$, $n_t | \lambda_t \sim \mbox{Poisson}(\delta_t)$ and $\mathbf{s}_{t,i} | \lambda_t, n_t \stackrel{iid}{\sim} f_t(\cdot)$ for $i \! \in \! \{1,\ldots,n_t\}$. 
As mentioned before, numerous studies have proposed sophisticated and accurate methods for estimating $\{\delta_t\}$. We thus focus on predicting the spatio-temporal demand density $\{f_t\}$, which has received far less attention in the literature. With the predicted $\{f_t\}$, we can furthermore predict spatio-temporal demand volume by multiplying our $\{f_t\}$ with $\{\delta_t\}$ from studies such as \cite{Matteson:2011}.

\subsection{Spatio-Temporal Kernel Density Estimation}
Suppose we observe and utilize historical ambulance demand from a set of past time periods $\mathcal{T}_{obs}$, and we would like to forecast demand for a future time period $u$. We propose to predict $f_{u}$ using a spatio-temporal weighted kernel density estimator. We place a bivariate spatial kernel $K$ at the location of each past observation in $\mathcal{T}_{obs}$, and weight each kernel by the informativeness of the corresponding observation in predicting for the $u$th time period. Specifically, we have for $\mathbf{s}\in \mathcal{S}$
\begin{equation}\label{kde}
f_{u}(\mathbf{s}) =\frac{\sum_{t \in \mathcal{T}_{obs}} w(\mathbf{s}_{t,i}, u)\,\,K_{\boldsymbol{H}} (\mathbf{s}-\mathbf{s}_{t,i})}{\sum_{t \in \mathcal{T}_{obs}} w(\mathbf{s}_{t,i}, u)}.
\end{equation}
Here, $w(\mathbf{s}_{t,i}, u)$ is the informativeness weight function multiplied by the spatial kernel of the past observation $\mathbf{s}_{t,i}$. This weight function is defined in detail in \S 2.2. $K$ is the chosen bivariate spatial kernel with bivariate bandwidth $\boldsymbol{H}$, i.e., 
\[
K_{\boldsymbol{H}}(\mathbf{s}-\mathbf{s}_{t,i})= \frac{1}{\left\vert \boldsymbol{H}\right \vert} K\left(\boldsymbol{H}^{-1/2}(\mathbf{s}-\mathbf{s}_{t,i})\right).
\] 

\subsection{Weight Function}
The weight function $w$ aims to capture the utility of a past observation in predicting demand at a future period. Specifically, we would like to incorporate in $w$ the spatial and temporal dependencies in the demand. Domain knowledge on EMS demand densities focuses our attention on weekly and daily seasonalities and short-term serial dependence of a few hours, which have varying strengths in different neighborhoods. 

We can therefore discretize the spatial domain into $C$ large spatial cells, representing a rough division into neighborhoods. We assume that temporal dependencies within each cell remain constant in space. Let $w_c$ denote the weight function local to each discretized cell $c\in \{1,\ldots,C\}$. Within this cell, we further assume that the informativeness of a past observation from time $t$ in predicting for future time $u$ only depends on how far back $t$ is from $u$. We use the weight function to measure how positively correlated two demand densities $(u- t)$ periods apart are in each cell. We  model the weight function as
\begin{equation}\label{w_c}
w_c(u-t) = \rho_{1,c}^{u-t} + \rho_{2,c}^{u-t}\rho_{3,c}^{\sin^2\!\left(\!\frac{\pi(u-t)}{T_1}\!\right)\!} \rho_{4,c}^{\sin^2\!\left(\!\frac{\pi(u-t)}{T_2}\!\right)\!},
\end{equation}
for $c\in\{1,\ldots, C\}$. We combine all $w_c$ to have 
\begin{equation}
w(\mathbf{s}_{t,i}, u) = \sum_{c=1}^C w_c(u-t)\,\mathds{1}_{\{\mathbf{s}_{t,i}\in \,\mbox{cell}\, c\}}.
\label{weight}
\end{equation}
Here, $\{\rho_{1,c}\}, \{\rho_{2,c}\}, \{\rho_{3,c}\}$ and $\{\rho_{4,c}\}$ for $c\in \{1,\ldots,C\}$ are all constrained to take values in $[0,1]$. We use a separate $\rho$ parameter to capture each typical EMS patterns for easy interpretation and comparisons across locations (e.g., downtown vs suburbs) and times (e.g., winter vs summer). The term $\rho_{1,c}^{u-t}$ describes any potential short-term serial dependence. Its parametric form is the same as a stationary first-order autoregressive model, AR(1), and is also equivalent to the squared exponential function often used in Gaussian processes \cite{Rasmussen:2006}. The term with $\rho_{3,c}$ describes any potential daily seasonality with $T_1 = 24$, whereas the term with $\rho_{4,c}$ describes any potential weekly seasonality with $T_2= 24\times 7=168$. The parametric form of these two terms corresponds to the periodic function used in Gaussian processes \cite{Rasmussen:2006}. These two seasonality terms are multiplied, and further discounted by a squared exponential function, $\rho_{2,c}^{u-t}$. Finally, we sum the short-term dependency effect and the seasonality effects. The different $\rho$ terms are combined in similar to the typical approach to combining covariance functions in Gaussian processes. . There may be other weight functions that work similarly; we draw inspirations from Gaussian processes because these functions are well-studied and have some nice properties (e.g., infinite differentiability). This parametrization of the weight function is easy to interpret and visualize, and flexible to experiment with, even for non-experts. 

The weight function is bounded between 0 and 2. We avoid negative weights to avoid negative kernels in the kernel density estimator, which complicates bandwidth selection, results in negative density estimates that need to be floored at zero, and produces discontinuities in the derivatives of the estimates \cite{Scott:1992}. The magnitudes of the weights are nominal, as long as they are comparable across all $C$ regions, since they are normalized in Equation (\ref{kde}).

Equations (\ref{kde}), (\ref{w_c}) and (\ref{weight}) together form the model.

\subsection{Computational Methods}

We must select or estimate the kernel function $K$ and bivariate bandwidth $\boldsymbol{H}$ in Equation (\ref{kde}), as well as the spatial discretization $C$ and $4C$ number of $\rho$ parameters in the weight function (\ref{w_c}). Since the nature of ambulance demand does not change drastically over time, these estimations may be performed infrequently in practice (at most several times a year). 

For $K$, we can use the typical Gaussian kernel, or for additional computational savings, the Epanechnikov kernel with bounded support. We can select the bandwidth $\boldsymbol{H}$ via the plug-in method \cite{Wand:1994} or smoothed cross-validation \cite{Duong:2005}. We can also adopt one of many fast computational methods for KDE, including kd-trees \cite{Bentley:1975}, ball trees \cite{Omohundro:1989}, dual trees \cite{Gray:2003} and statistical regular pavings \cite{Sainudiin:2013}.

For the weight function (\ref{w_c}), we can choose the discretization mesh or $C$ \textit{a priori} or via cross validation. A larger value of $C$ allows personalized temporal patterns on a finer grid, but if $C$ is too large, data may become too sparse for accurate estimation of temporal dependencies. In our application, $C$ is best chosen to be close to 20, yielding discrete regions that are roughly 5~km by 5 km each. The $4C$ number of $\rho$ parameters in the weight function could be chosen in a number of standard ways; for instance we could use stochastic gradient ascent to maximize the joint likelihood of training data. For accurate estimation, we would need to use training data with tens of thousands of observations, and incur non-trivial computational cost. Here we introduce a much faster alternative method to estimate these parameters.

In Equation (\ref{w_c}), $w_c$ measures how positively correlated two demand densities $(u- t)$ periods apart are at cell $c$. We can directly estimate this correlation as follows. For each cell $c$, we can approximate its demand density for any period by the proportion of observations arising from this cell out of all observations from that period. We can then obtain a time series of proportions and compute its (discrete) autocorrelation function $A_c(\ell)$  for lag $\ell\in \{1,\ldots, L\}$, where $L$ is the maximum lag considered. Typically $L$ can be taken to be around several weeks (hundreds of one-hour periods). The non-negative part of this autocorrelation, $A_c^+(\ell)$, or a smoothed version of it, is precisely what $w_c$ aims to capture. For example, Figure 2 (a) shows an example of the autocorrelation function $A_c(\ell)$, and the grey lines in Figure 2 (b) shows the corresponding $A_c^+ (\ell)$, for $\ell\in \{1, \ldots, 672\}$ (up to 4 weeks of one-hour periods). 

The goal is to find appropriate $\rho$ parameters such that $w_c$ best fits the shape of $A_c^+$. To do this, we would like to minimize the sum of squared errors between $\rho_{0,c} w_c(\ell)$ and $A_c^+(\ell)$ at all time lags $\ell$ from 1 to $L$. We can find the optimal $\rho_{0,c}$ to $\rho_{4,c}$ for this minimization using gradient descent or grid search. The extra parameter $\rho_{0,c}$ is needed to scale $w_c$ to curve-match the magnitude of $A_c^+$, and is of no real significance. Of greater importance is curvature or shape of $A_c^+$, which is captured in $\rho_{1,c}$ to $\rho_{4,c}$. To make $w_c$ comparable across all $C$ cells, we need to normalize $w_c$ such that the area under $w_c$ up to $L$ is the same across different cells. 

In summary, we estimate the $\rho$ parameters in $C$ minimization problems: for each $c\in \{1,\ldots,C\}$,
\begin{align} 
&\min_{\rho_{j,c}, \forall \,j\in\{0,\ldots, 4\}} \,\,\sum_{\ell=1}^{L} \left( A_c^+(\ell)-\rho_{0,c}w_c(\ell) \right)^2  \label{rho_est} \\
&\mbox{s.t.} \,\,  \sum_{\ell=1}^{L} \, w_c(\ell) = 1. \nonumber
\end{align}

This computation is much more efficient than the joint estimation of $4C$ parameters by maximizing likelihood. Here, we do not need to involve the kernel density estimator, nor loop through tens of thousands of ambulance demand observations. We can easily compute the $C$ minimization problems in parallel. For each cell, we have a low-dimensional (5 parameters) problem with a small number of observations $L$ (around hundreds of hours of time lags). A wide array of standard algorithms for solving optimization problems can be applied. For example, we can Lagrangian relax the constraint into the objective and use the genetic algorithm or particle swarm. 

Once the infrequently performed parameter estimation is done, predictions for any future time period can be calculated instantaneously using short sliding windows of length $L$. We can additionally refine or customize the prediction procedure in the following two ways.

First, to boost predictive accuracy, we can bilinearly interpolate the weight values smoothly over the spatial domain instead of taking only $C$ sets of values on a discretized grid. This is appropriate since we believe that the temporal patterns vary smoothly across the spatial domain. It also mitigates the sensitivity to predictions induced by choices of $C$. 

Secondly, we can impose an omission threshold value, $\mathcal{O}$, for the weights. If the weight of a past observation $\boldsymbol{s}_{t,i}$ is below this threshold, i.e., if $w(\boldsymbol{s}_{t,i},u)<\mathcal{O}$, we can omit this observation in the calculation of weighted KDE by overriding $w(\boldsymbol{s}_{t,i},u)=0$. The threshold can be chosen to balance the tradeoff between computational expense and predictive accuracy.

\section{Predicting Toronto Ambulance Demand}
The computation has two stages. In the first stage, we estimate or choose all parameters, including the kernel $K$, bandwidth $\boldsymbol{H}$, discretization $C$ and $4C$ number of $\rho$ parameters. This estimation only needs to be performed infrequently. For this parameter estimation, we use Toronto ambulance data from January to July 2008. Figure 1 shows the spatial locations of all observations from this 7-month period. In the second stage, we predict future ambulance demand on a sliding window of length $L=672$ (4 weeks, around $15,000$ observations) for each one-hour period from August to December 2008. 

In estimation, we choose the Gaussian kernel for $K$, select the bandwidth $\boldsymbol{H}$ via the plug-in method \cite{Wand:1994} and discretize Toronto into $C=21$ equally-sized regions. We estimate the $\rho$ parameters in the weight function using the method detailed in \S 3. As an example, we outline the cell $c$ covering downtown Toronto in Figure 1. We show in Figure 2 (a) the autocorrelation function $A_c$ for the proportions of observations arising from this downtown cell out of all observations across hourly time periods. This autocorrelation function indicates weekly, daily seasonalities and low-order serial dependence. Figure 2 (b) shows in grey $A_c^+$ and in black the fitted weight function $\rho_{0,c} w_c$ for downtown, with $\rho_{1,c}=0.95$ (short-term serial dependence), $\rho_{3,c}=0.001$ (daily seasonality), $\rho_{4,c} =0.145$ (weekly seasonality) and $\rho_{2,c}=0.9995$ (discounting for seasonalities). The fitted weight function provides an interpretable basis for understanding exactly which historical observations are the most important for prediction. For example, from Figure 2 (b), an EMS manager can recognize that at downtown, ambulance demand in the past day or two and corresponding hour of the past few weeks are the most important. Cross-correlations among the 21 weight functions estimated at different regions show that neighboring weight functions have some association, but those far apart are not correlated.

\begin{figure}[!ht]
\centering
\includegraphics[width=\linewidth,height=2.3in]{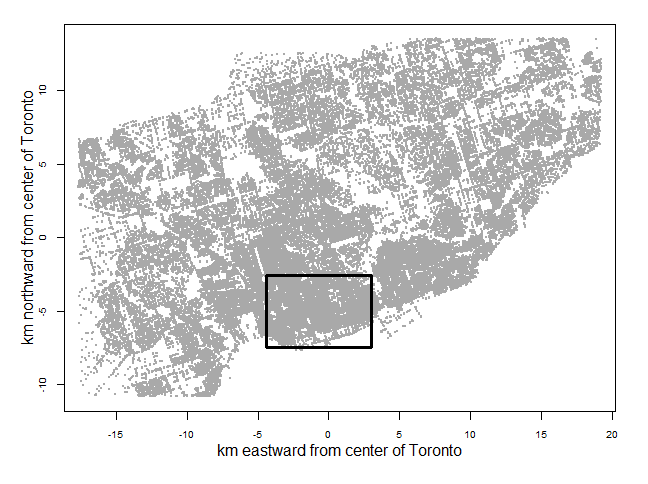}
\caption{Spatial locations of all Toronto ambulance demand data from January to July 2008. To evaluate location-specific weight functions, we discretize the spatial domain into 21 cells, and here we outline the cell containing downtown Toronto.}
\end{figure}

\begin{figure}[!ht]
\centering
\includegraphics[width=\linewidth]{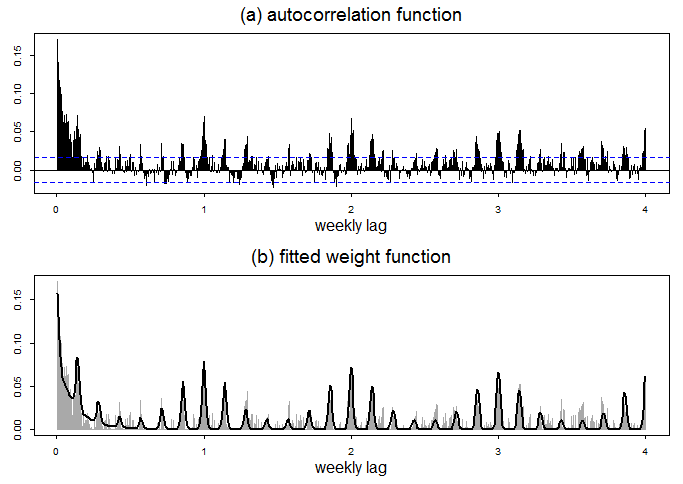}
\caption{(a) The autocorrelation function of the proportions of observations arising from the rectangle in Figure 1 over all observations across one-hour periods. (b) The fitted weight function (black) against the nonnegative part of the autocorrelation function (gray).}
\end{figure}

Once parameter estimation is done, we predict forward using a sliding window of 4 weeks for each one-hour period from August to December 2008. Figure 3 shows the predictive densities on August 6, 2008 (Wednesday) at two different time periods. The ambulance demand is, not surprisingly, concentrated at the heart of downtown during day time on Wednesday (Figure 3 (b)), and more spread out throughout the city during night time on Wednesday (Figure 3 (a)). This illustrates that stKDE can differentiate temporal patterns at different time periods and locations.

\begin{figure}[!ht]
\centering
\includegraphics[width=\linewidth,height=4.9in]{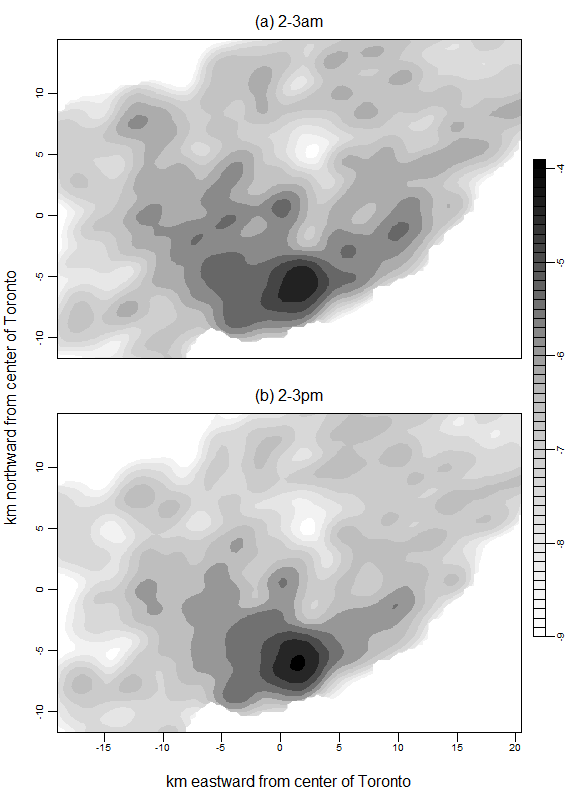}
\caption{Log predictive density using stKDE for Aug 6, 2008 (Wednesday) at (a) 2 - 3 am (demand more spread out at night) and (b) 2 - 3 pm (demand concentrated at downtown during the day).}
\end{figure}

We compare stKDE to the following competing methods
\begin{enumerate}[label=(\alph*)]
 \item The MEDIC method, which is an industry practice implemented in Charlotte-Mecklenburg, NC (\S 1). We implement this method as far as we have data. The cell count in a 1-km$^2$ region and a 1-hour period is predicted by the average of corresponding cell counts in the preceding 4 weeks in the past two years. The log predictive density produced by MEDIC for August 6, 2008 (Wednesday) at 2 - 3 am is shown in Figure 4. Compared to Figure 3 (a), the MEDIC prediction appears much noisier.
 \item Two naive KDEs, (i) using data from the most recent hour to predict the next hour, and (ii) using all data from the past four weeks with equal weights (this produces a spatial only model, with almost no temporal variation). 
 \item A time-varying Gaussian mixture model. We quote results from \cite{Zhou:2015a} implemented on Toronto data with different training / testing months and various modeling specifications (e.g., number of components). The computational expense is considerable. 
\end{enumerate}

\begin{figure}[!ht]
\centering
\includegraphics[width=\linewidth, height=2.3in]{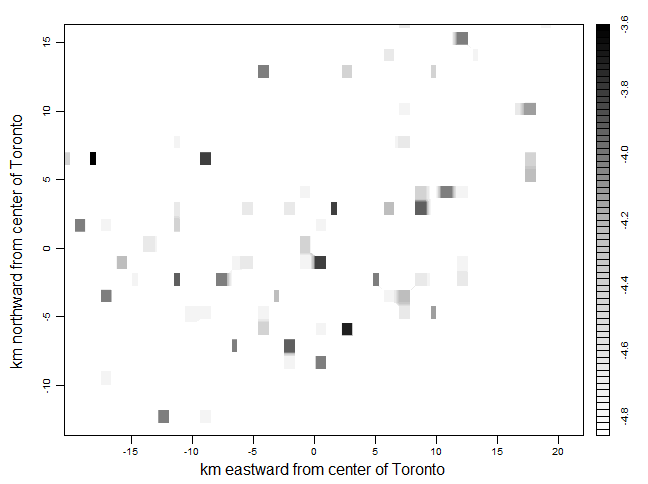}
\caption{Log predictive density using industry method for Aug 6, 2008 (Wednesday) at 2 - 3 am. Figure 3 (a) shows a less noisy density for the same period predicted by stKDE.}
\end{figure}

To compare the statistical predictive accuracies of our model and the comparison methods, we use the metric of average log score \cite{Good:1952}. This performance measure is widely used because it is a strictly proper scoring rule closely related to Bayes factor and Bayes information criterion \cite{Gneiting:2007}. It is the average log likelihood of test data, and is defined as 
\[
\mbox{Accuracy\!}  \left(\{\tilde{\mathbf{s}}_{u,i}\}\right)=\frac{1}{\sum_{u\in \mathcal{T}_{test}}n_t}\sum_{u\in \mathcal{T}_{test}}\sum_{i=1}^{n_t}\log \hat f_u(\tilde{\mathbf{s}}_{u,i}),
\]
in which $\mathcal{T}_{test}$ are the test time periods, $\{\tilde{\mathbf{s}}_{u,i}\}$ are the test data, and $\hat f_u(\cdot)$ is the predictive density for the $u$-th period obtained either by various methods. Intuitively, it measures the probability that we observe test data given a model. Thus a less negative (higher) average log score indicates that the model is better at capturing the test data. 

We show in Table 1 the predictive accuracies produced by stKDE and the comparison methods. We present three variations of stKDE prediction: (i) using the estimated discretized weight functions $w_c$ as they are, (ii) spatially interpolating the estimated weight values, and (iii) imposing an omission threshold on the estimated weight values such that each prediction uses a comparable amount of data as the industry method (about 200 observations).

The stKDE method significantly outperforms the MEDIC method (industry practice). It also outperforms the naive KDE methods, demonstrating the utility of incorporating spatio-temporal patterns via the weight functions. Our performance is comparable to time-varying GMM as it is implemented on Toronto data with different training / testing months and modeling specifications. Among the three variations of stKDE, allowing for bilinear interpolation of weight values improves the predictive accuracy slightly. In the third variation, including the omission threshold leads to a small loss of accuracy but reduces computational cost significantly to be comparable to the industry method.

\begin{table} 
\centering 
\begin{tabular}{l l c} 
\hline
\multicolumn{2}{c}{\textbf{Prediction method}}  & \textbf{Accuracy}   \T\B \\ 
\hline
stKDE &  & $-6.106$ \T \\   
&  + interpolation & $-6.102$\\ 
&  + threshold (less data)  & $-6.635$  \B \\ 
\hline
MEDIC & & $-8.642$ \T \B \\  
naiveKDE &  most recent hour & $-6.921$ \T \\
&   all equal weights & $-6.254$ \B \\ 
GMM & & $-6.072$ to $-6.149$ \T \B \\ 
\hline
\end{tabular}
\caption{Predictive accuracies of stKDE and competing methods. Results of GMM are quoted from \cite{Zhou:2015a} implemented on Toronto data with various training / testing months and model specifications.} 
\label{table:kde}
\end{table}

The infrequent estimation of weight functions and bandwidth takes several hours on a personal computer. This offline training is significantly shorter than that of GMM (inferring latent component labels in GMM is costly). It does not take much longer than estimating bandwidths for naive KDE methods. Once estimation is done, making each new prediction is instantaneous (a few seconds). We could further reduce the computational expense of stKDE by parallelizing weight estimation, using a tree-based algorithm for fast KDE computation, using a bounded kernel function, or creating a look-up table of densities (none of these was done).

\section{Conclusions}
Fine-resolution spatio-temporal ambulance demand predictions are crucial to optimal ambulance planning. The EMS industry practice and early studies either sacrifice predictive accuracy for fast computation, or incur substantial computational cost in pursuit of high accuracy. We provide a much-needed prediction method that is both accurate and fast. We predict spatio-temporal ambulance demand in Toronto with higher accuracy and comparable computational cost as a typical industry practice. 

We propose a spatio-temporal weighted kernel density estimator. The spatial kernel of each historical observation is multiplied with a weight value to indicate the informativeness of this historical observation to the current predictive task. The spatio-temporal weight functions are inferred from dependencies in data, are unique to each neighborhood and can be updated regularly. This is an improvement from the ad hoc heuristic that only accounts for the weekly and yearly seasonality across the entire city. The weight functions are also flexible to represent various spatial and temporal characteristics. They are easy to experiment with, visualize and interpret by non-experts. Moreover, stKDE easily handles missing data by placing zero weight and scaling up weights on other data proportionally. It can also easily predict many hours or days into the future. 

The proposed method provides efficient estimation of the weight function, and offers customizable prediction to balance the tradeoffs between accuracy and computational cost. It is straightforward to implement by non-specialized users and scalable to large-scale datasets, and can be easily generalized to a wide range of other applications with spatial-temporal point process data. 

\section{Acknowledgments}
The authors sincerely thank Toronto EMS for sharing their data, and support from a Xerox Faculty Research Award and NSF grant DMS-1455172.

\bibliographystyle{abbrv}
\bibliography{kdebib}  

\end{document}